\documentclass[12pt]{iopart}

\usepackage{graphicx}
\usepackage{authblk}
\usepackage{color}
\usepackage{wasysym}
\usepackage{csvsimple}
\usepackage{booktabs}
\usepackage{lineno}
\modulolinenumbers[5]

\def\UVA{University of Virginia, Charlottesville, Virginia 22904, USA}
\def\Shooltz{Shooltz Solutions LLC,  Lansing, MI 48912 ,USA}
\def\UCLA{University of California, Los Angeles, Los Angeles, California 90024,USA}

\begin{document}

\title[Fiber Performance for the Mu2e CRV]{Performance of Wavelength-Shifting Fibers for the Mu2e Cosmic Ray Veto Detector}

\author{}{E.~C.~Dukes, P. J.~Farris, R.~C.~Group, T. Lam\footnote{Present address~\UCLA}, Y.~Oksuzian}
\address{\UVA}
\ead{rcg6p@virginia.edu}

\author{}{D. Shooltz}
\address{\Shooltz}

\begin{abstract}
The cosmic-ray-veto detector (CRV) for the Mu2e experiment consists of four layers of plastic scintillating counters read out by silicon photo-multipliers (SiPM) through wavelength-shifting fibers.  This paper reports the light properties of several wavelength-shifting fiber samples with diameters of 1.0 mm, 1.4 mm, and 1.8 mm that were considered for the CRV system. A fiber diameter of 1.4~mm was selected as optimal for the CRV, and measurements of the prototype and production fiber of this diameter are presented.  Fiber performance was found to exceed the CRV requirements for $>$99\% of the spools.  The measurements are performed using a scanner designed to ensure the fiber quality for the CRV.

\end{abstract}



\maketitle

\interfootnotelinepenalty=10000

\section{Introduction}
The Mu2e experiment \cite{tdr,Abusalma:2018xem} intends to search for neutrino-less conversion of a muon into an electron in the presence of a
nucleus. If the Standard Model (SM) is expanded to include neutrino masses then this process will occur, but at a rate approximately thirty orders of
magnitude below the current best experimental limits. However, many SM
extensions suggest that this process will occur at an enhanced rate
that could be observable experimentally~\cite{Bernstein:2013hba}. An observation of this conversion would be an
unambiguous sign of physics beyond the SM. 

The total background
at Mu2e is estimated to be less than 0.4 events after three years of data
taking. However, without a veto system approximately one background event will be produced per day at Mu2e due to cosmic-ray muons. In order to achieve the
required sensitivity, Mu2e needs to suppress the cosmic-ray background
by four orders of magnitude. The cosmic-ray-veto (CRV) system has been
designed to cover the Mu2e apparatus and veto cosmic-ray backgrounds
with 99.99\% efficiency. The CRV consists of plastic scintillator
counters up to 7~m long that are read out through wavelength-shifting (WLS) fibers. The CRV efficiency critically depends on the performance of the WLS fibers. A WLS fiber scanner was designed in order to measure the light properties and ensure the quality of the fiber used in the CRV system.

This paper describes the fiber scanner and how it was used to study the 1.0 mm, 1.4 mm, and 1.8 mm diameter fibers considered for use for the CRV. Spectral attenuation measurements for the fiber and other studies are presented.  These studies were used to ensure the quality of the large fiber order (60 km) used to fabricate the CRV.

\section{Wavelength-shifting fiber}
The Mu2e experiment will use Kuraray~\cite{kuraray} Y11 WLS fibers
for the CRV system. The fluorescent dye, K27, in the Kuraray Y11
fibers absorbs the blue light (375-475 nm) from the
scintillating counters and re-emits the light in the green (450-600
nm) spectral region. The fibers are read
out by $2.0{\times}2.0$\,mm$^2$ (model S13360-2050VE, pixel size of 50\,$\mu$m) Hamamatsu silicon photomultipliers (SiPMs)~\cite{sipm}. The spectral photon
detection efficiency response of the chosen SiPM type
is well matched to the emission spectrum of Kuraray Y11 fiber. We have
selected multi-clad and non-S type fiber due to the enhanced light yield
and attenuation length properties that it provides~\cite{kuraray}.

\section{Fiber scanner}
\begin{figure}[htb]
\centering 
\includegraphics[height=2.5in]{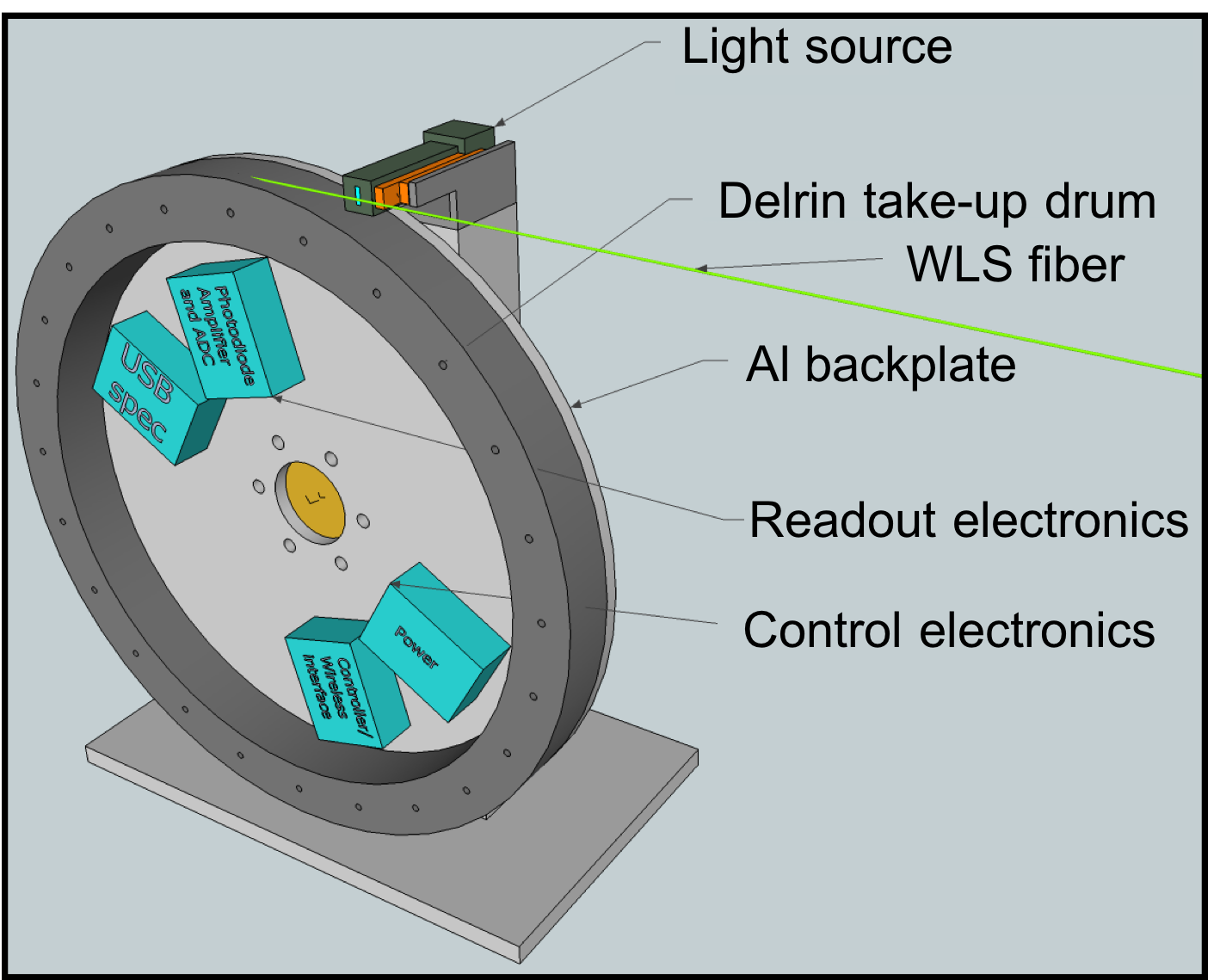}
\includegraphics[height=2.5in]{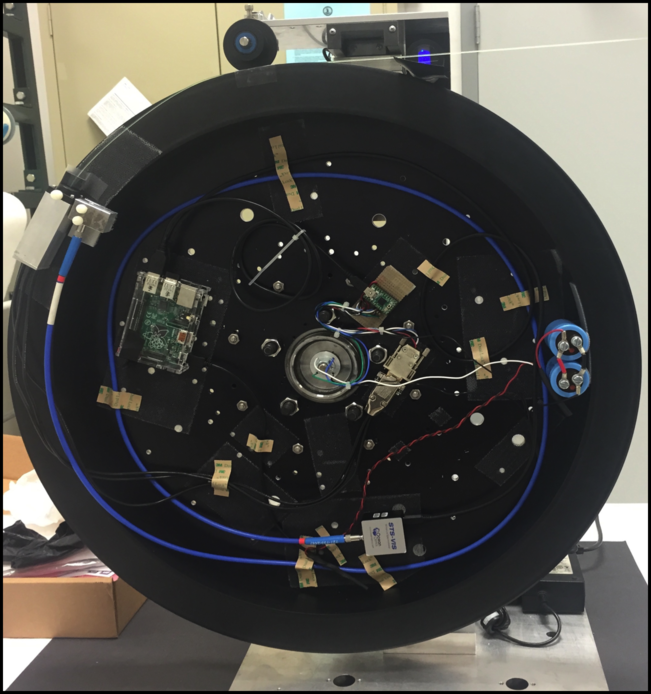}
\caption{A schematic of the fiber scanner is shown on the left, while a photo is shown on the right.}
\label{fig:scanner}
\end{figure}

A fiber scanner was designed and produced in order to study the fiber light yield as a function of wavelength and distance from the light source (Fig.~\ref{fig:scanner}).  The fiber performance is used to assure the quality of the fibers used for the CRV system. The fiber scanner consists of a large diameter (62 cm)
take-up drum carrying the optical readout hardware. The diameter
of the take-up drum was selected to accommodate up to 25 m of the large-diameter (1-2 mm) WLS fibers that were considered for the CRV system. The radius of the take-up drum is larger than the minimum bending radius specified by the vendor for this type of fiber~\cite{kuraray}.  The device was designed to accommodate different fiber diameters as the fiber size for the CRV had not been selected at that time.   

A blue LED light source~\cite{led} excites the WLS fiber and is read out by a
large-area Hamamatsu S1227-1010BR photodiode \cite{photodiode} or a STS-VIS
USB Ocean Optics spectrophotometer \cite{sts}, depending on the type of measurement desired. The take-up drum is
driven by a stepping motor and controlled for gentle acceleration and
deceleration. The fiber is delivered by the manufacturer on 90 cm diameter cardboard spools containing about 0.5 km of fiber. When measurements are made, the fiber remains on the original spool and the fiber end is epoxied into a ferrule and polished with a diamond-bit flycutter before it is connected to either the photodiode or the spectrometer. The drum can increment at any predetermined distance, stopping for measurements of the light spectra and intensity at different distances from the readout device. The fiber scan procedure is controlled remotely via WiFi through a web interface on a Raspberry Pi~\cite{ref:pi} mounted on the take-up drum. During data taking the Raspberry Pi controls the stepper motor and collects and stores the data from the photodiode or the spectrometer.

 The large photodiode area relative to the fiber diameter provides measurements which are not very sensitive to effects from fiber misalignment or diameter differences. The photodiode has a uniform efficiency \cite{photodiode} in the spectral region of the WLS fiber emission. However, the spectral response of the SiPM~\cite{sipm} used in the CRV system features a peak sensitivity at 450 nm. Therefore, the light attenuation measured in the CRV system using SiPMs will differ slightly from the one measured with the photodiode. Even though the light attenuation measurements cannot be directly compared, a photodiode scan can identify compromised fiber with sharp drops in the light yield or poor attenuation.  For studies of attenuation, the light yield measurements at 50 cm increments over 25 m of fiber were obtained. The short (long) attenuation components are  extracted by making a single-exponential fit to the measurements taken at 0.5~m to 3~m (3~m to 25~m) from the readout end.

The spectrometer provides spectral measurements of the light emission from the fiber with a high spectral resolution and high signal-to-noise ratio in a wide (350-800 nm) spectral region. The spectral response is also obtained at 
points in 50 cm increments over the first 25 m of fiber. For each wavelength, a single attenuation length is extracted using an exponential fit of the spectral light intensity as a function of distance from the readout end.  A single exponential function is found to yield a reasonable fit over the full 0.5 m - 25 m range when applied to single spectral measurements.  

\section{Results}

\subsection{Comparisons of fiber diameter}

\begin{figure}[htb]
\centering
\includegraphics[height=3.0in]{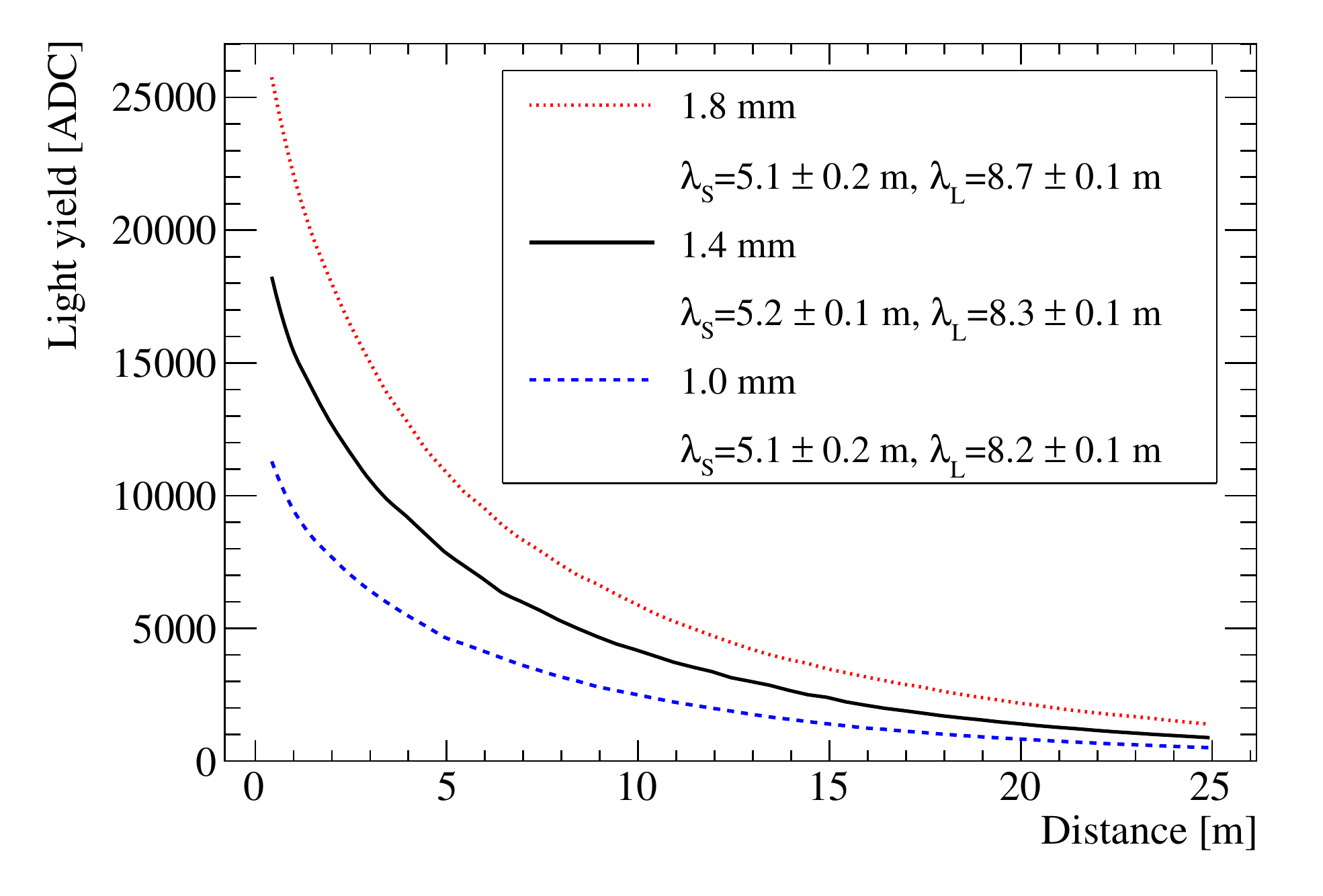}
\caption{Light yield measurements for 1.0, 1.4 and 1.8 mm Kuraray Y11
  WLS fiber, using the fiber scanner and the photodiode readout scheme.}
\label{fig:diodeatten}
\end{figure}

  Three fiber diameters were originally considered for the CRV design: 1.0, 1.4
and 1.8 mm~\cite{proc}.  The result of the photodiode scan for all three diameters is presented in Fig.~\ref{fig:diodeatten}.  The light yield for 1.4 and 1.8 mm fiber is higher than the light yield from 1.0 mm fiber by approximately constant factors of 1.7 and 2.3, respectively.

  Counters with 1.0 mm, 1.4 mm, and 1.8 mm fiber were fabricated and evaluated in the Fermilab Test Beam Facility~\cite{TestBeam2016} using a 120 GeV proton beam.  The 1.4~mm diameter fiber was selected for use in the CRV based on the test beam results and requirements that were derived from detailed simulations of the CRV performance~\cite{tdr}.

\subsection{Pre-production fiber}

\begin{figure}[htb]
\centering
\includegraphics[height=1.9in]{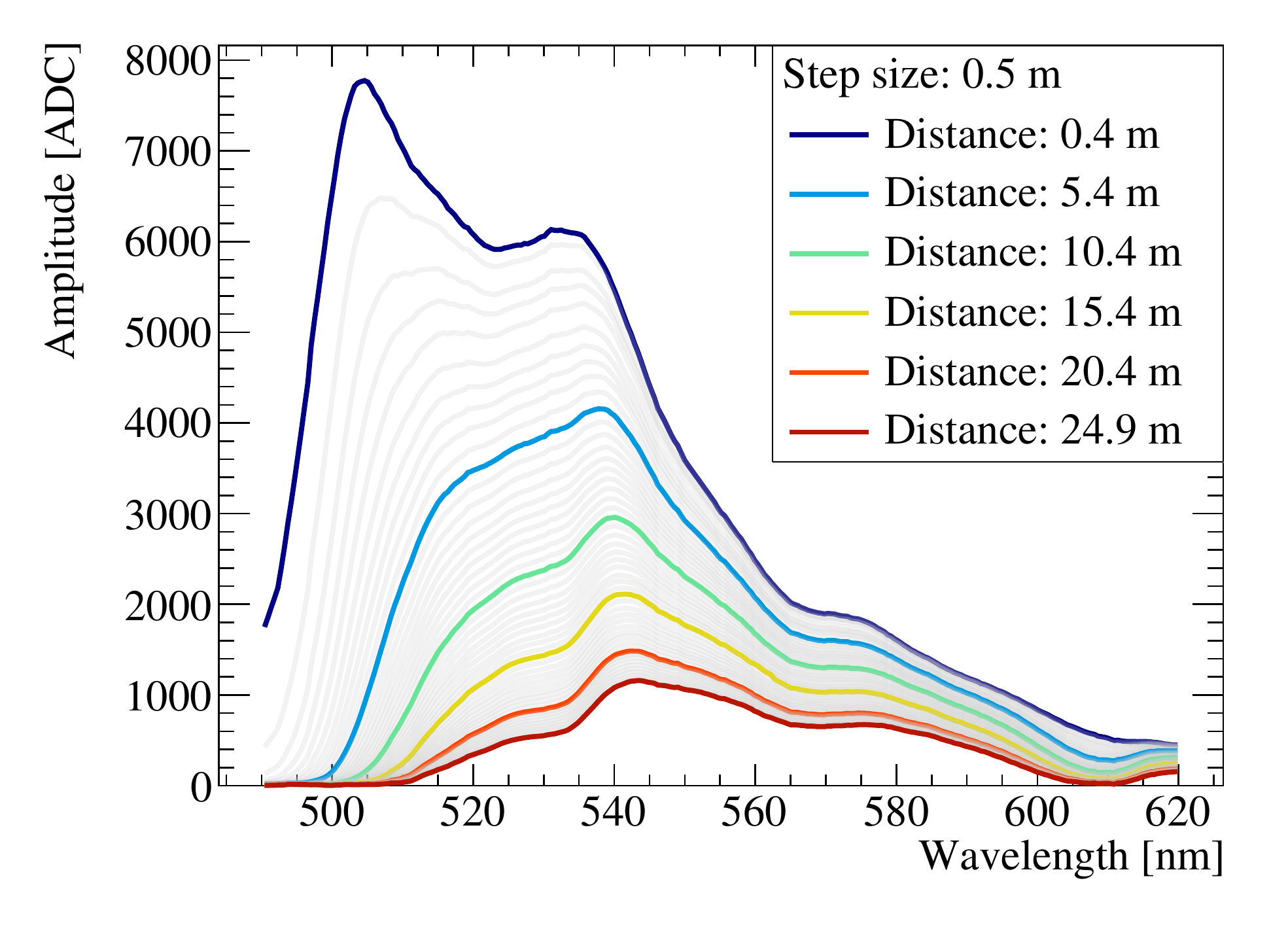}
\includegraphics[height=1.9in]{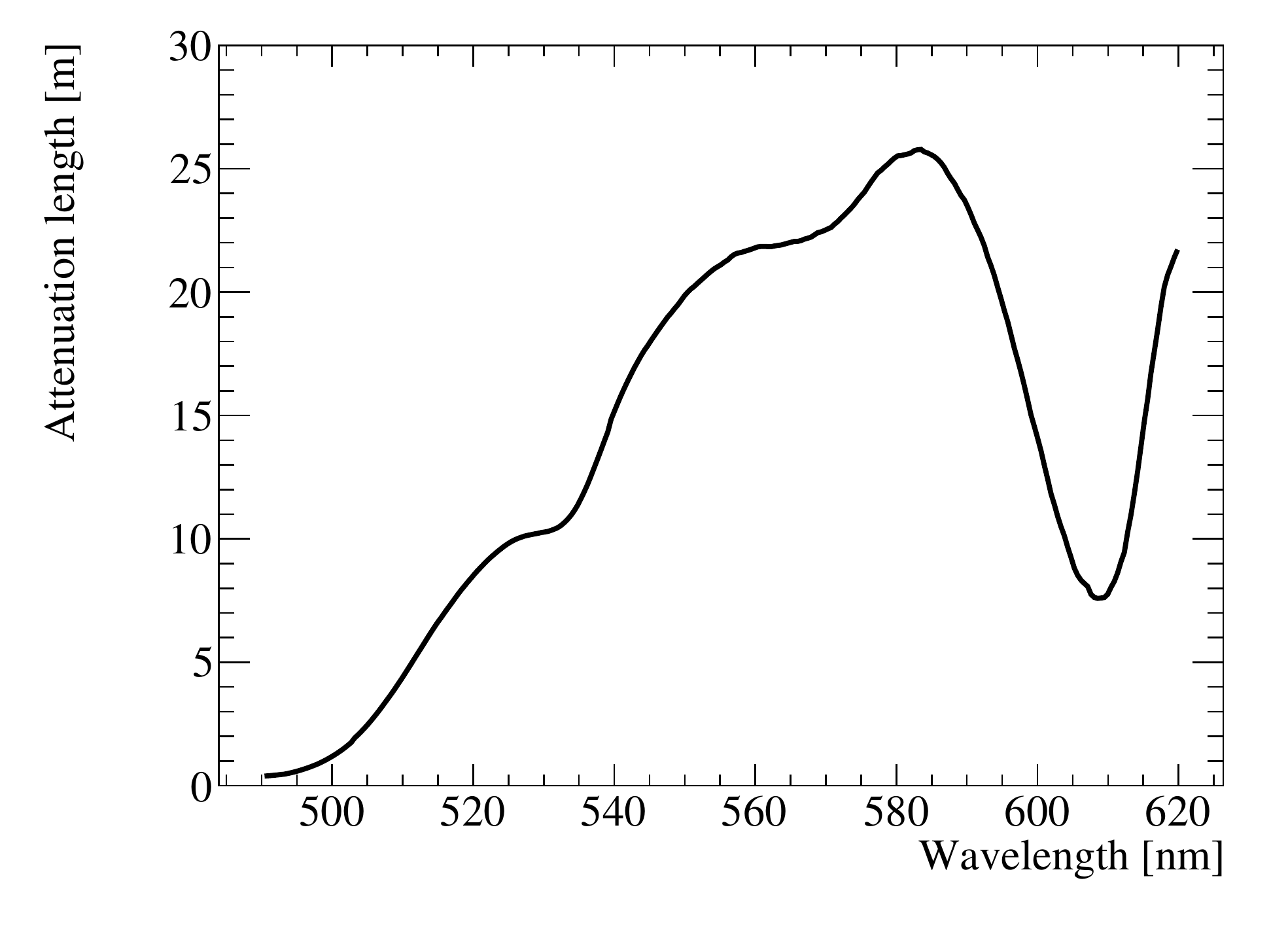}
\caption{Left: Spectrometer readings as a function of fiber emission wavelength for the reference spool. The individual curves correspond to measurements taken at various distances between fiber excitation points and the spectrometer readout.  Right: Light attenuation length as a function of fiber emission wavelength as determined from a single-exponential fit for each wavelength value. }
\label{fig:spectromreads}
\end{figure}

 Approximately 1 km of 1.4 mm fiber was purchased to produce the CRV pre-production prototype detectors.  Response for these prototypes was also measured in the test beam and found to meet requirements~\cite{TestBeam2016}.  One spool of 1.4 mm pre-production fiber, whose performance was measured in the test beam, was identified as a benchmark spool against which all production fiber could be compared.  

The result from a  spectrometer scan of the 1.4~mm fiber reference spool, as shown in
Fig.~\ref{fig:spectromreads}, suggests that the shorter
wavelength spectrum is attenuated at a significantly higher rate. Figure~\ref{fig:spectromreads}(right) shows the attenuation length values as determined by a single exponential fit for various wavelengths. Kuraray Y11 WLS fiber emission peaks near 500~nm, but there is still a significant component for the 520 - 600 nm wavelength where a long ($>10$ m) light attenuation is obtained~\cite{kuraray}.  Because of this property, photodetectors with high sensitivity in this spectral region are clearly preferable for long scintillation detectors like the CRV.  While the SiPMs used in the CRV system (model S13360-2050VE) features a peak sensitivity at 450 nm, they provide a broad spectral sensitivity that is still above 80\% (50\%) of the peak sensitivity at 550 nm (620 nm)~\cite{sipm}.

\subsection{Production fiber}

\begin{figure}[htb]
\centering
\includegraphics[width=3.0in]{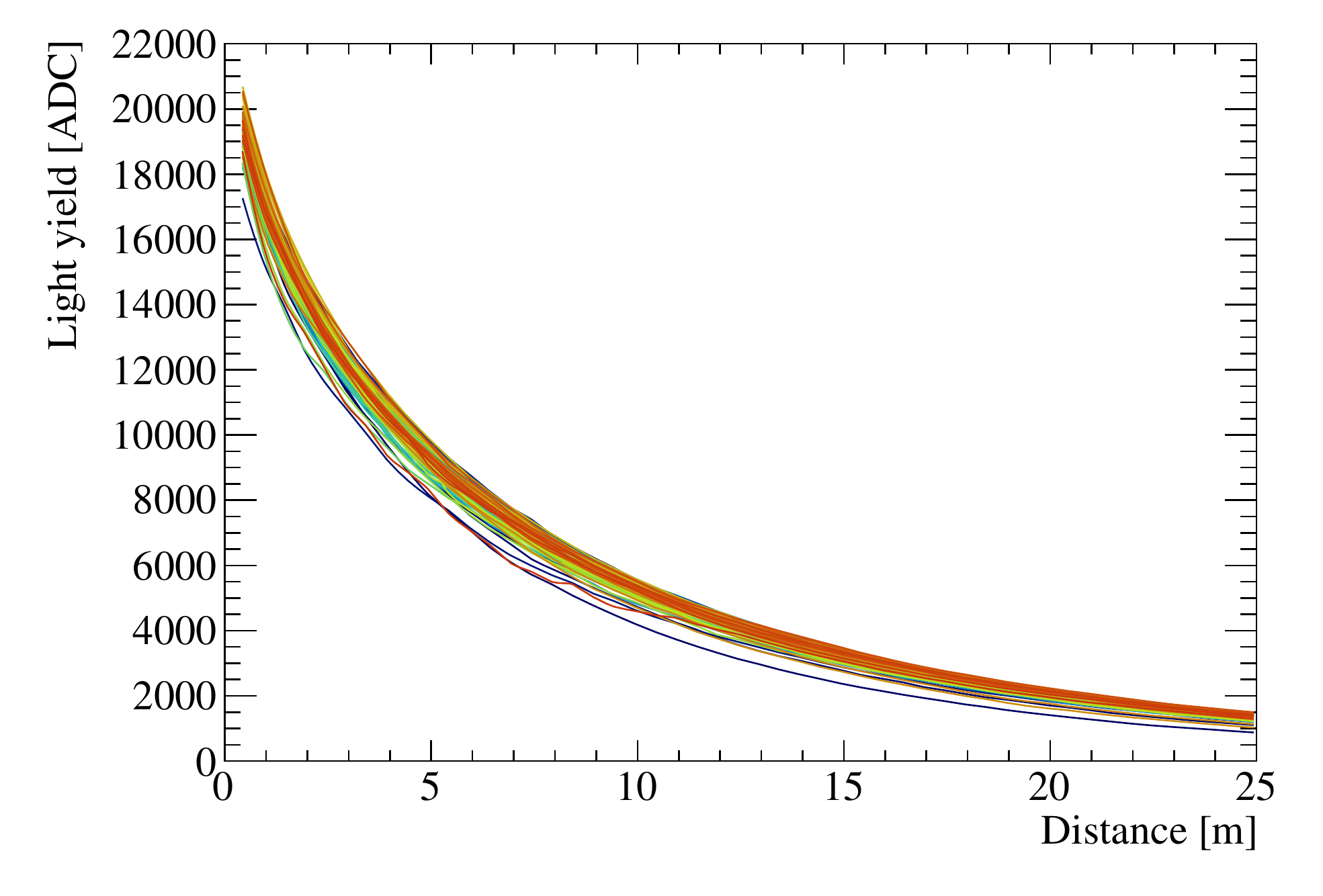}
\includegraphics[width=3.0in]{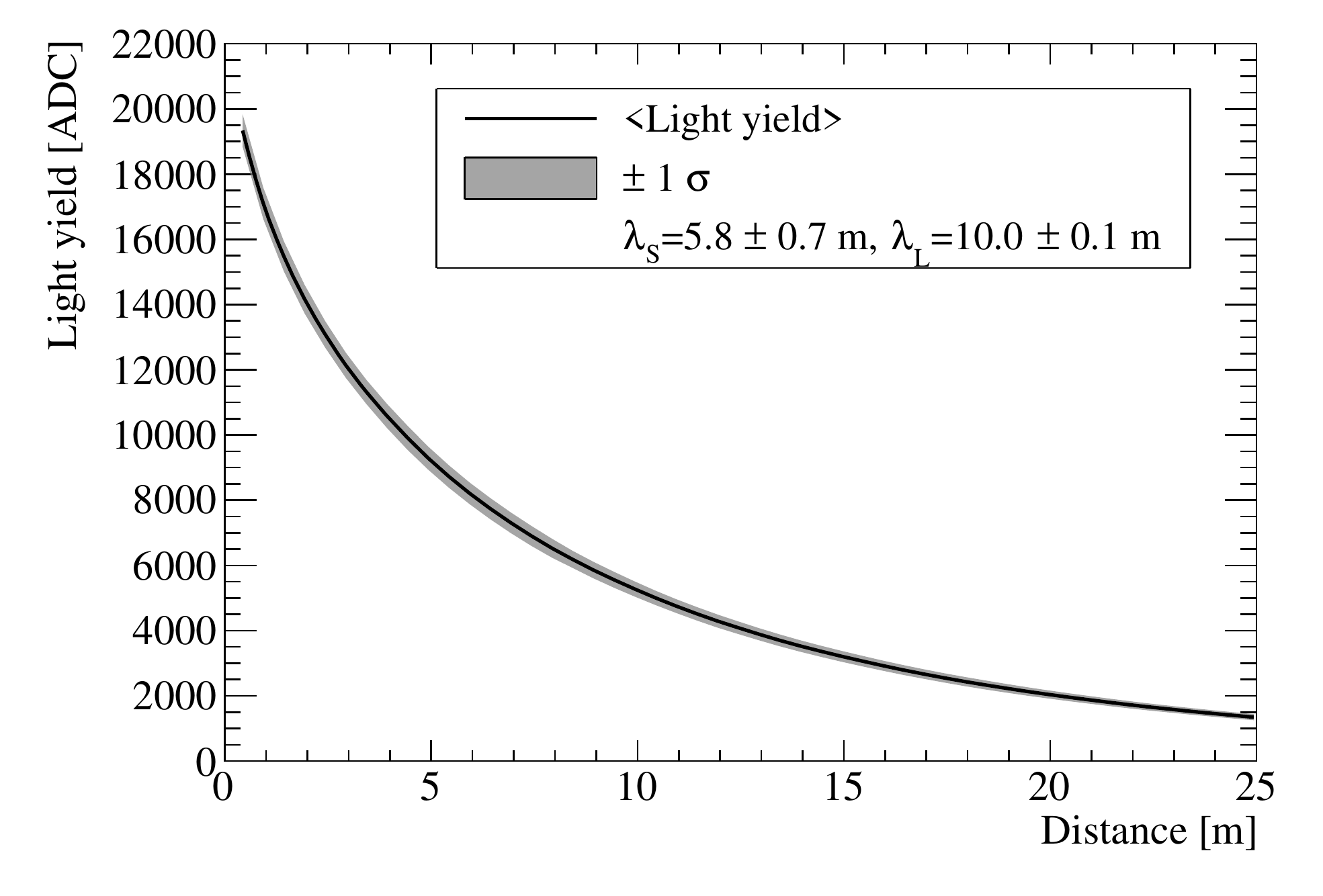} 
\caption{Light attenuation for all production spools (left) and the average (right).}
\label{fig:ProductionYield}
\end{figure}

\begin{figure}[htb]
\centering
\includegraphics[height=2.3in]{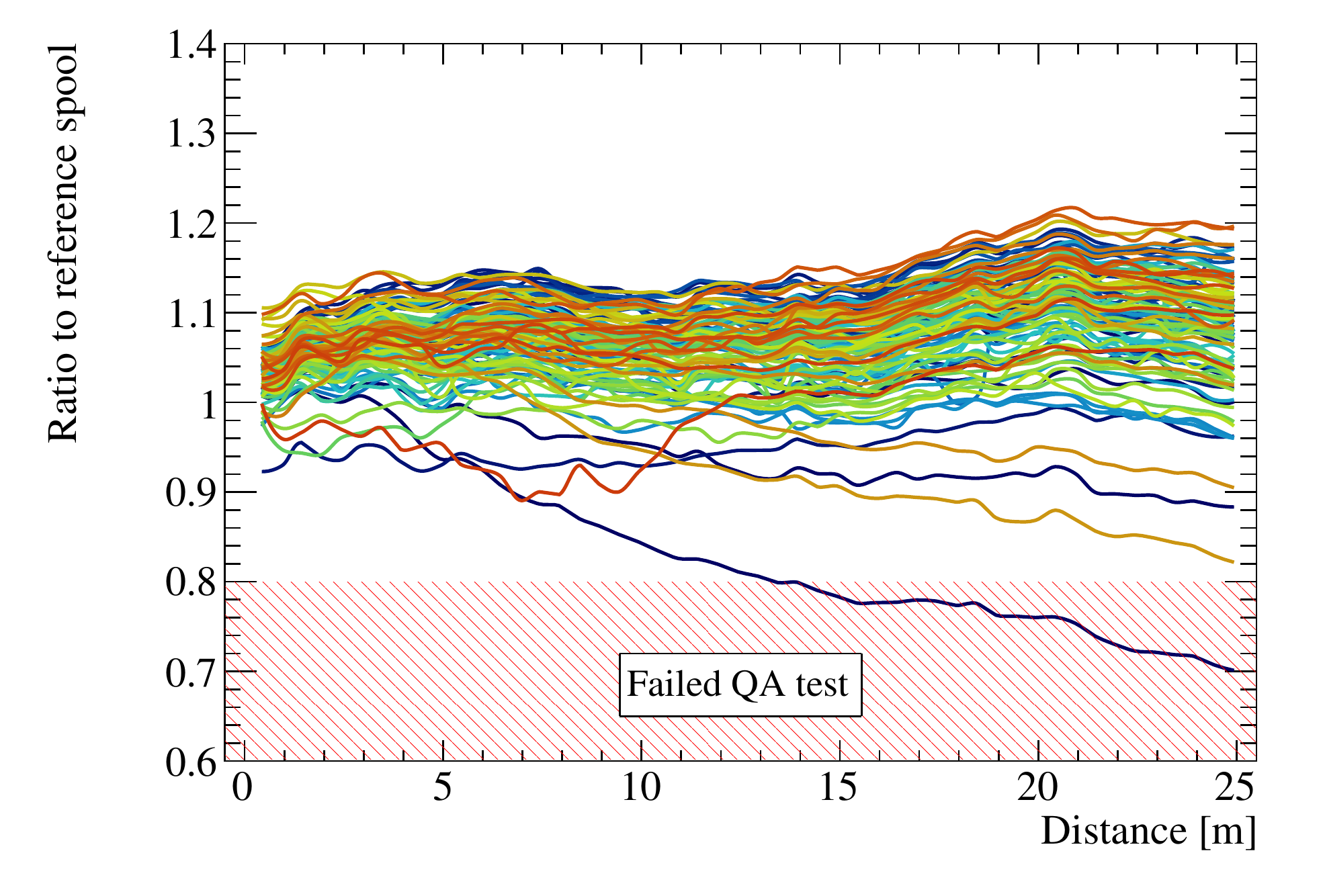}
\caption{The light yield as a function of the distance from the readout end of the LED source for 104 production spools shown as a ratio to the benchmark spool.}
\label{fig:ProductionYieldRatio}
\end{figure}

\begin{figure}[htb]
\centering
\includegraphics[width=3.0in]{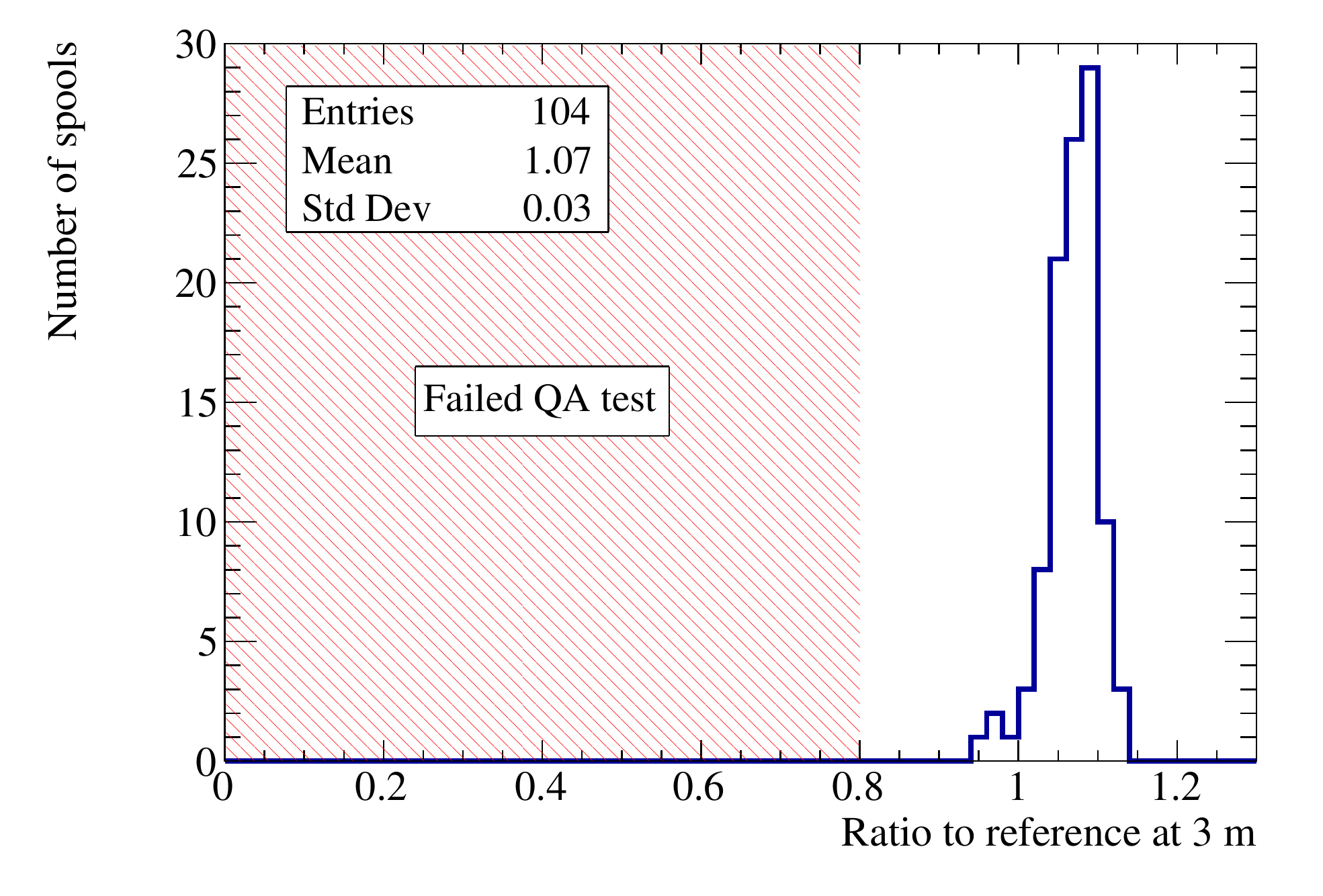}
\includegraphics[width=3.0in]{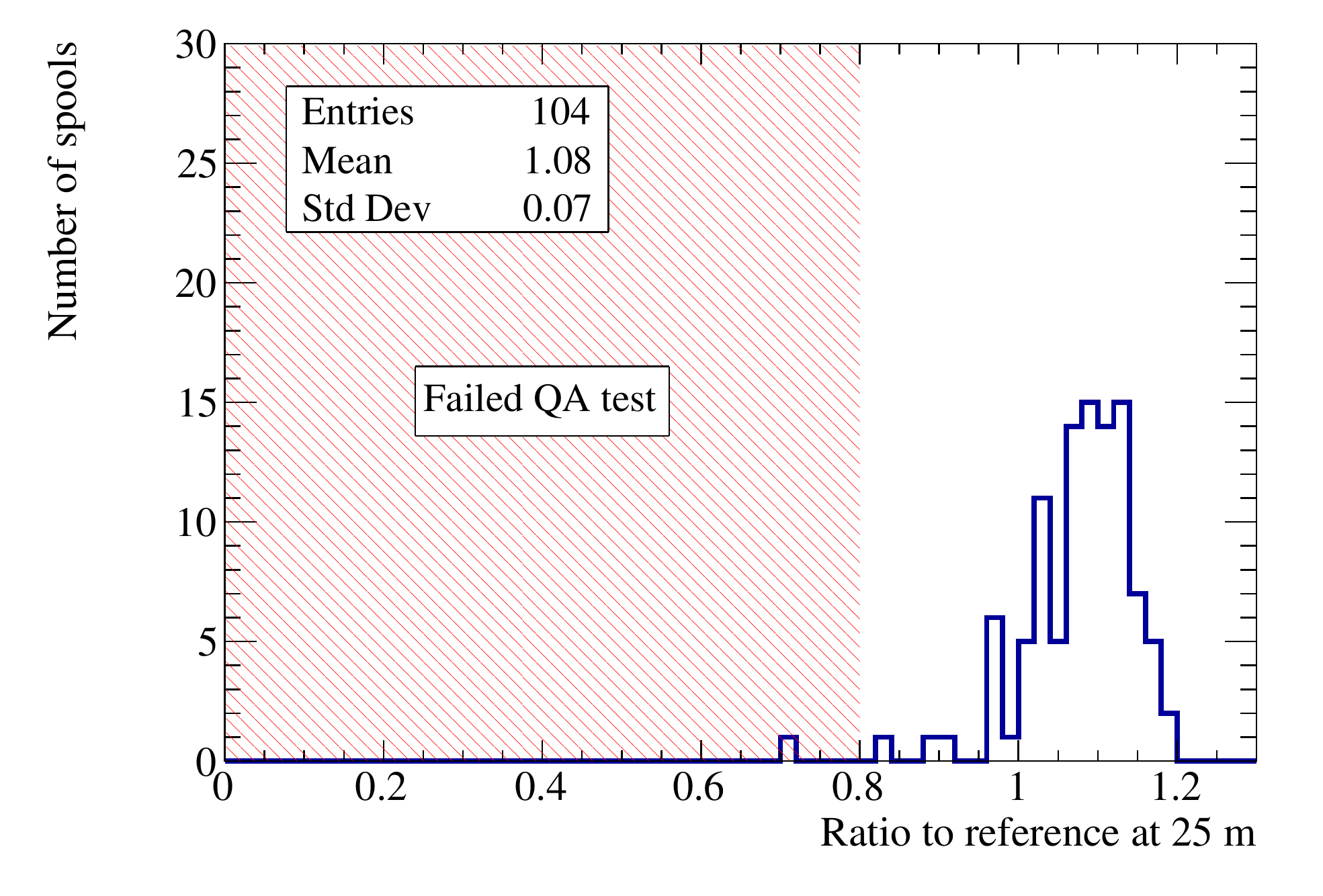}
\caption{The distribution of fiber response as a ratio to the reference spool for the 104 production spools with the LED source at 3m (left) and 25m (right) from the readout end.}
\label{fig:Ratio}
\end{figure}

 	The CRV detector is composed of about 5500 counters and almost 60 km of wavelength-shifting fiber.  The production fiber was delivered on 104 spools in May of 2018 and all of the spools were tested with the fiber scanner by the end of June 2018\footnote{The fiber diameters were also measured for all 104 spools as part of the quality control program.  A mean of 1.39 mm and sigma of 0.01 mm were obtained.}.  The light yield for the production spools is shown as a function of the the distance from the LED source in Fig~\ref{fig:ProductionYield}.  It is shown normalized to the benchmark spool in Fig.~\ref{fig:ProductionYieldRatio} (the response for each spool divided by the response of the benchmark spool).  A distribution of the fiber response for 3~m and 25~m from the readout end is shown in Fig.~\ref{fig:Ratio}.   On average, the response for the production fiber is about eight percent better than the benchmark spool.
 

\begin{figure}[htb]
\centering
\includegraphics[height=2.3in]{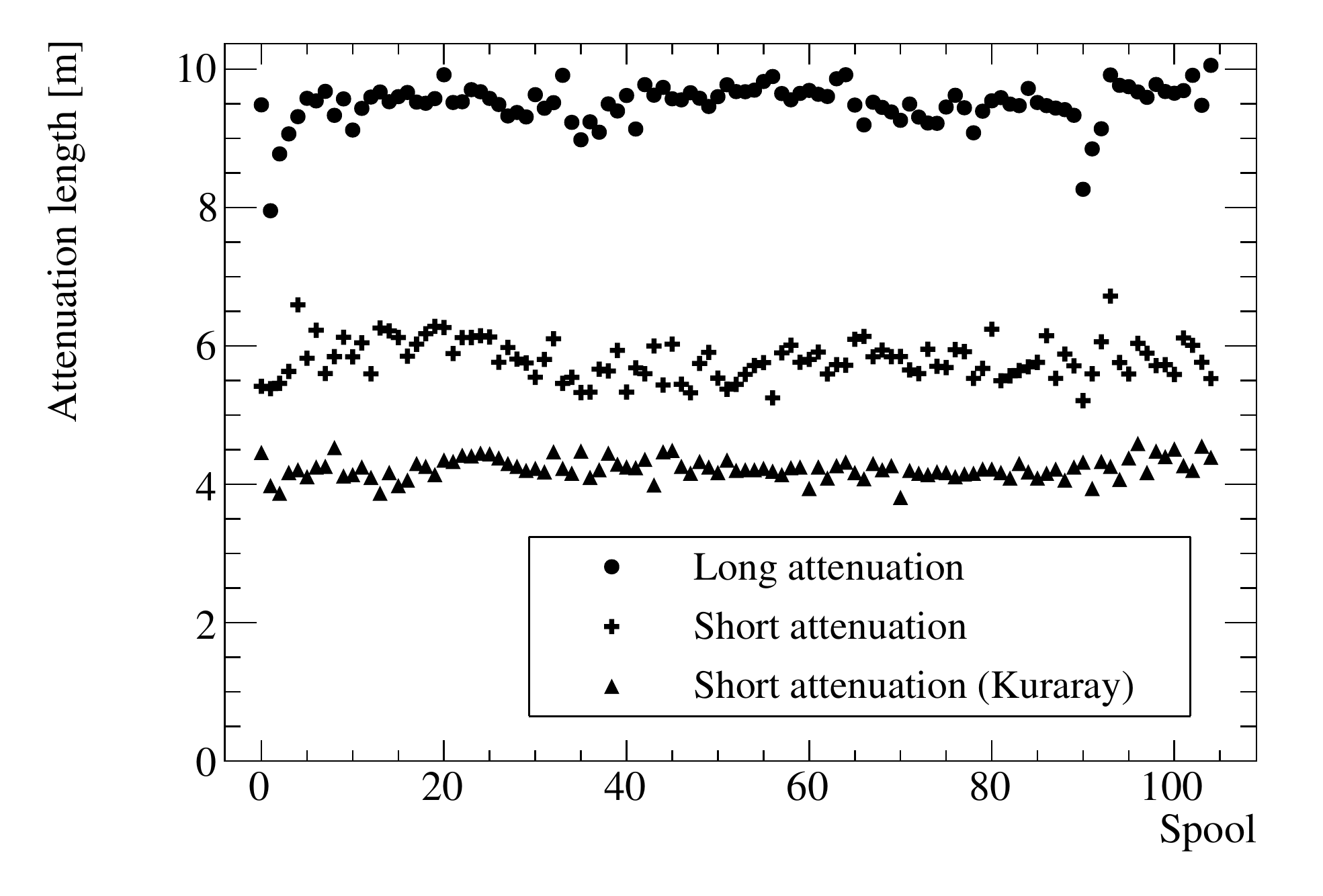}
\caption{The attenuation lengths as a function of time measured by Kuraray in the region of 1-3 m and Mu2e in the region of 0.5-3~m (short) and 3-25~m (long).  The first point is the reference spool.  The following points represent the production spools in order of their production date.  The long attenuation clearly improves for the first few spools.}
\label{fig:timeline}
\end{figure}

      The light yield at each distance, measured by the photodiode,  is required to be 80\% or greater than the light yield of the benchmark spool.  In both Fig.~\ref{fig:ProductionYieldRatio} and Fig.~\ref{fig:Ratio} there are several spools with a significantly lower response than the benchmark spool.  Two of these correspond to the first two production spools produced by the vendor.  
In Fig.~\ref{fig:timeline} the short (0.5-3~m) and long (3 - 25~m) attenuation length obtained from an exponential fit is shown for each fiber spool.  The first point represents the benchmark spool while the other points are sorted based on spool production order.  A trend of improving long attenuation is clearly visible in the first few spools produced by the vendor. A similar trend starting from the 90th spool is also visible.
      
\begin{figure}[htb]
\centering
\includegraphics[width=3.0in]{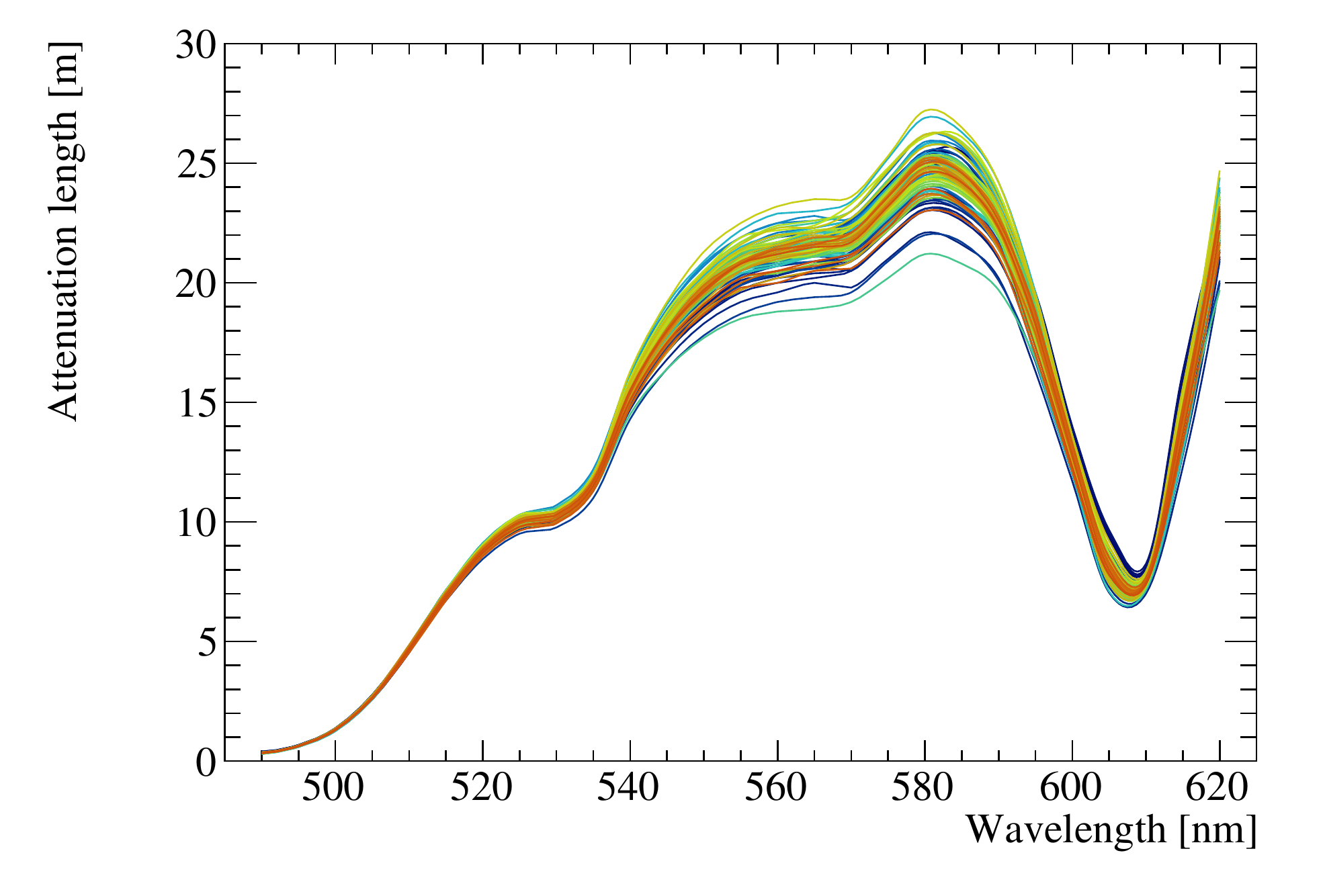}
\includegraphics[width=3.0in]{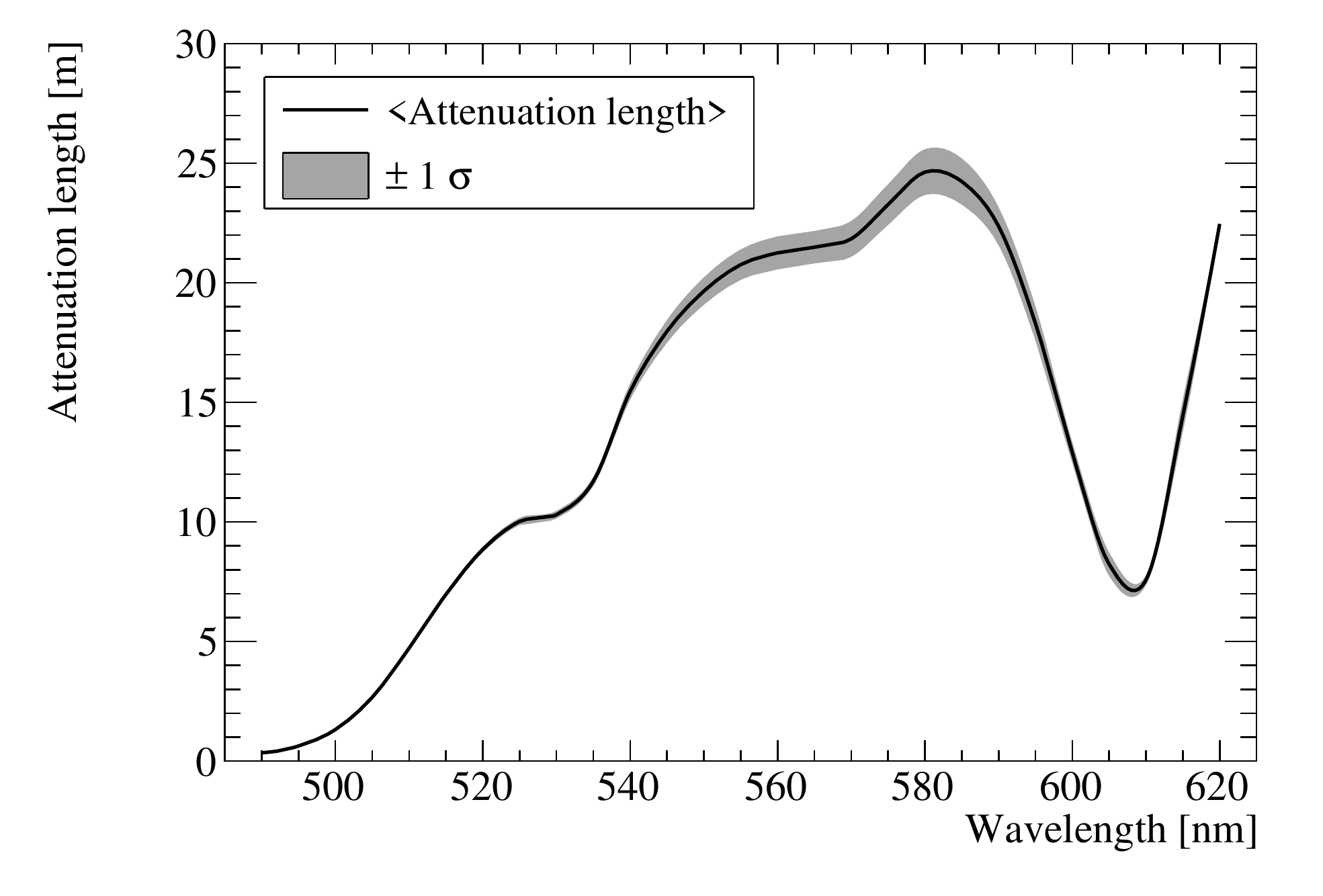} 
\caption{Light attenuation length as a function of fiber emission wavelength for all production spools (left) and the average of all spools (right). The attenuation length is extracted using a single exponential fit to the spectrometer data taken from the first 25~m of each fiber spool.}
\label{fig:ProductionSpectrum}
\end{figure}

      The attenuation length measurements using the spectrometer are shown in Figure ~\ref{fig:ProductionSpectrum}.  Only one spool, the first in the production batch, failed QA requirements, and it was promptly replaced by the vendor.

\renewcommand{\thefootnote}{$\star$} 

\subsection{Fiber quality control at Kuraray}
In addition to the quality assurance steps performed by the Mu2e collaboration, Kuraray performed a set of fiber quality acceptance requirements as a part of its quality control procedure. The measurements were shared by the vendor. Kuraray measured the fiber diameter and eccentricity at $1400.3 \pm 8~{\mu}m$ and 0.2 $\pm$ 0.1\%, respectively. The attenuation length in the range of 1-3 m was measured to be 4.23 $\pm$ 0.15 m, well above the promised value of 3.6 m for each fiber spool (Fig.~\ref{fig:timeline})~\footnote{Kuraray uses a 3~m fiber sample with a bialkali PMT and blue LED(445nm) to make their measurements~\cite{kuraray}.  Note that the attenuation length measured by the CRV fiber scanner is not expected to match the Kuraray result due to differences in the spectral sensitivities on the photo-detectors used.}. The standard deviation of the light yield intensity measurements at 2.85 m from the readout end was required to be within 15\% of the mean, and was measured to be within 8\%.

\section{Conclusion}
We report the measurements of the light yield and attenuation length of
Kuraray Y11 WLS fibers. We performed the studies using a
fiber scanner designed for the CRV system of the Mu2e experiment. The
fiber scanner features a fast and reliable data acquisition system,
and it was an essential component for fiber selection and fiber quality assurance during the production phase of the CRV detector.   The fiber scanner
was used to study the properties of fibers of various diameters. The
results show a significant light yield gain from larger diameter
fibers. Kuraray Y11 WLS fibers yield long ($>5$ m) light attenuation
values for the 510 - 600 nm wavelength spectrum.  The fiber quality assurance measurements are also presented for the production spools of the Mu2e CRV detector.  Fiber performance was measured to be reasonably consistent across the production order and performance exceeds the CRV requirements for $>$99\% of the measured spools.

\section{Acknowledgements}

We are grateful for the vital contributions of the Fermilab staff and the technical staff of the participating institutions.
This work was supported by the US Department of Energy; 
the Istituto Nazionale di Fisica Nucleare, Italy;
the Science and Technology Facilities Council, UK;
the Ministry of Education and Science, Russian Federation;
the National Science Foundation, USA; 
the Thousand Talents Plan, China;
the Helmholtz Association, Germany;
and the EU Horizon 2020 Research and Innovation Program under the Marie Sklodowska-Curie Grant Agreement No.690835. 
This document was prepared by members of the Mu2e Collaboration using the resources of the Fermi National Accelerator Laboratory (Fermilab), a U.S. Department of Energy, Office of Science, HEP User Facility. Fermilab is managed by Fermi Research Alliance, LLC (FRA), acting under Contract No. DE-AC02-07CH11359.  In addition, the authors would like to acknowledge Carl Bromberg for numerous discussions about fiber and his willingness to share his experiences from managing the fiber for the NOvA experiment.  Finally, the authors would like to thank Kuraray and their staff for providing the QA measurements and allowing us to discuss them in this paper.  



\end{document}